\documentclass[aps,prl,twocolumn,showpacs,a4paper]{revtex4}
\usepackage{graphicx}

\begin{document}





\title{Ground State and Tkachenko Modes of a Rapidly Rotating Bose--Einstein Condensate in 
the Lowest Landau Level State}
\author{E.B. Sonin}

\affiliation{Racah Institute of Physics, Hebrew University of Jerusalem, Jerusalem 91904,
Israel   }

\date{\today}

\begin{abstract} 
The Letter considers the ground state and the Tkachenko modes for a
rapidly rotating Bose-Einstein condensate (BEC), when its macroscopic wave function is
a coherent superposition of states analogous to the lowest Landau levels of a charge in
a magnetic field. As well as in type II superconductors close to the critical magnetic
field $H_{c2}$, this corresponds to a periodic vortex lattice. The exact value
of the shear elastic modulus of the vortex lattice, which was known from the old works on
type II superconductors, essentially exceeds the values calculated recently for BEC. This
is important for comparison with observation of the Tkachenko mode in the rapidly
rotating BEC.
\end{abstract}



%



\maketitle

A rapidly rotating Bose-Einstein condensate (BEC) of cold atoms is now a subject of
intensive experimental and theoretical investigations
\cite{exp,Cod,SHM,Ba,Ho,WBP,GB,CKR,Dal}.  It is well known that rotation gives rise to a
regular triangular vortex lattice. At moderate rotation speed this is a lattice of vortex
lines (points in the 2D case) with the core size of the order of coherence length $\xi$,
which essentially smaller than the intervortex distance $b= \sqrt{\kappa/\sqrt{3}\Omega}$.
Here $\Omega$ is the angular velocity of rotation and  $\kappa=h/m$ is the circulation
quantum. One can call it the Vortex Line Lattice (VLL) regime.  With increasing $\Omega$,
the vortex lattice becomes more and more dense and eventually enters the regime, in which
the vortex cores start to overlap, i.e., 
$\xi$ becomes larger than the $b$. This regime is analogous to the mixed state of a type-II
superconductor close to the second critical magnetic field $H_{c2} \sim \Phi_0/\xi^2$
($\Phi_0$ is the magnetic flux quantum), at which the transition to the normal state takes
place. However, in a rotating BEC there is no phase transition at corresponding
``critical'' angular velocity $\Omega_{c2}
\sim \kappa/\xi^2$. Instead the crossover to the new regime takes place: At $\Omega \gg
\Omega_{c2}$ all atoms condensate in a state, which is a coherent superposition of
single-particle states similar to the Lowest Landau Levels (LLL) of a charged particle in a
magnetic field (the LLL regime). An important method of investigation of the vortex
structure is studying its collective modes. Coddington {\em et al.} \cite{exp} were able to
detect the Tkachenko modes (transverse sound in the vortex lattice) in the VLL regime
experimentally. Recently they increased the rotation speed in the attempt to reach the LLL
regime  \cite{Cod}. They revealed softer  Tkachenko modes as was predicted by the theory
for the LLL regime \cite{SHM,Ba}. 

Theoretical study of the Tkachenko mode requires good knowledge of the equilibrium
state. A number of papers addressed this issue using the analogy with the quantum Hall
effect \cite{Ho,Ba,WBP,GB,CKR,Dal}. They started from the LLL wave functions for
noninteracting particles in a trapping potential and switched interaction on  after it.
For a regular vortex lattice this yielded the Gaussian density profile \cite{Ho} but it
was unstable with respect to small distortions of the lattice, which transformed the
Gaussian profile to the inverted parabola (Thomas-Fermi distribution)
\cite{Ba,WBP,GB,CKR,Dal}. This Letter suggests another strategy. One can
start from an infinite periodical vortex lattice in an infinite uniform liquid, neglecting
first the trapping potential but taking into account interaction.  The exact wave function
for this state was found in the classical work by Abrikosov \cite{Abr} for type II
superconductors close to
$H_{c2}$, and later it was generalized for an arbitrary unit cell of the vortex lattice
\cite{Kl,SG}. If a trapping potential is added, the vortex lattice (as well as the
liquid density) ceases to be uniform. But as far as modulation by the trapping potential
is smooth (the cloud size is much larger than the intervortex distance) it does not
distort the lattice essentially (except for the extreme periphery of the cloud) and one
can use the thermodynamic potential derived for a uniform vortex lattice. The suggested
approach is especially useful for investigation of the Tkachenko mode since for the infinite
uniform lattice the shear elastic modulus can be calculated exactly and for the vortex lattice
in type II superconductors it was done many years ago \cite{sup}. The previous calculations of
the shear modulus in BEC \cite{SHM,Ba} yielded the values smaller than the exact one. This
difference is important for comparison with recent experiments on Tkachenko modes \cite{Cod}. 
.

We consider a 2D rotating BEC in a parabolic trapping potential characterized by the
frequency $\omega_\perp$.  In the Gross-Pitaevskii theory the Gibbs thermodynamic
potential is
\begin{eqnarray} 
G=-\mu |\psi|^2+{\hbar^2\over 2m}\left|\left(-i \nabla -{2\pi\over \kappa}
\vec v_0
\right)\psi\right|^2 +{g\over 2} |\psi|^4 \nonumber \\ +
{m(\omega_\perp^2-\Omega^2)r^2\over 2}|\psi|^2~.
         \label{G} \end{eqnarray} 
Here $\psi$ is the BEC wave function, $\mu$ is the chemical
potential, $g$ is the interaction constant, and $\vec v_0=[\vec \Omega \times \vec r]$ is
the velocity of solid body rotation. The Gibbs potential is invariant with respect to the
gauge transformation $\psi \rightarrow \psi e^{i\phi}$, $\vec v_0 \rightarrow \vec v_0
+(\kappa/2\pi)\vec \nabla \phi$. At rapid rotation the potential $m\Omega^2 r^2/2$ of
centrifugal forces nearly  compensates the trapping potential
$m\omega_\perp^2 r^2/2$, Though stability requires that $\omega_\perp >\Omega$, at the
first stage of the analysis one can assume that $\omega_\perp =\Omega$ and BEC is
infinite in size. Then the Gibbs potential Eq. (\ref{G}) is invariant with respect to
translation accompanied by the gauge transformation, which corresponds to the shift of
the rotation axis.

\begin{figure}
  \begin{center}
    \leavevmode
    \includegraphics[width=\linewidth]{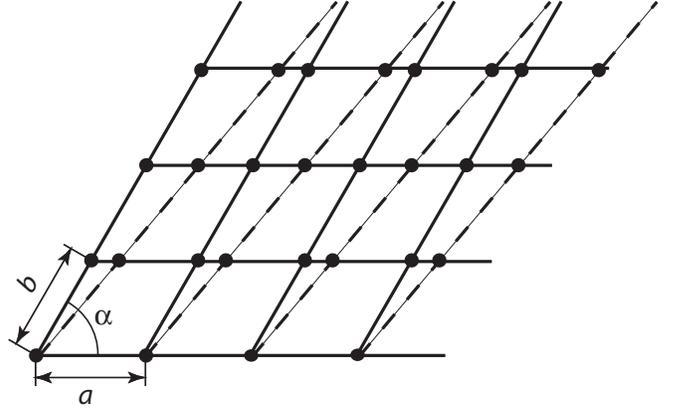}
    \caption{Vortex lattice before (solid lines) and after (dashed lines) shear deformation
}
  \label{fig1}
  \end{center}
  \end{figure}

As well as for type-II superconductors close to $H_{c2}$, in zero-order approximation one
can neglect interaction in the LLL regime $\xi \gg b$. Then the linearized Schr\"odinger
equation is similar to that for a charged particle in a uniform magnetic field: 
\begin{equation}
\mu\psi=-{\hbar^2\over 2 m} \left[\left({\partial \over
\partial x} -i{2\pi v_{0x}\over
\kappa} \right)^2 + \left({\partial \over
\partial y} -i{2\pi v_{0y}\over
\kappa} \right)^2\right]\psi~.
   \label{Sch}     \end{equation} At $\mu=\hbar \Omega$ it has a solution, which
corresponds to the lowest Landau level:
\begin{equation}
\psi_k \propto \exp\left[ikx-{(y-y_k)^2\over 2l^2}\right]~,
    \label{psik}    \end{equation} where $l^2=\kappa/4\pi\Omega$ and $y_k=-l^2k$. The
solution is given for the gauge with $\vec v_0(-2\Omega y, 0)$. The frequency $2\Omega$ is
the analog of the cyclotron frequency
$\omega _c =eH/mc$ for an electron in a magnetic field. If we consider a square $L\times L$
with periodic boundary conditions, then $k=-2\pi n /L$ with the integer $n$. Using the
condition  $0<y_k<L$, one can see that the integer $n$ should vary from zero to the integer
closest to $L^2 /2\pi l^2$. This is the total number of LLL states, and the density of the
LLL states is $1/2\pi l^2$. All these states are  orthogonal each other and have the same
energy. But degeneracy is lifted by taking into account the interaction energy. The
solution, which corresponds to the periodic vortex lattice with one quantum per lattice
unit cell,  is \cite{SG}
\begin{equation}
\psi_=\sum_n C_n\exp\left[inkx-{(y+l^2nk)^2\over 2l^2}\right]~,
        \end{equation} where $C_{n+1}=C_n \exp(2\pi ib \cos \alpha/a )$, $a$, $b$, and the
angle $\alpha$ are the parameters of the unit lattice cell (see Fig. \ref{fig1}). 

This solution yields the thermodynamic potential of the infinite BEC in the LLL regime:
\begin{equation} G=(-\mu +{\hbar\Omega}) n+{g\over 2} \beta n^2~,
      \label{GL} \end{equation} 
where $n=\langle |\psi|^2\rangle$ is the average
particle density and the parameter  \cite{SG}
\begin{eqnarray}
\beta={\langle |\psi|^4\rangle\over \langle
|\psi|^2\rangle^2}=\sqrt{\sigma}\left\{\left|\theta_3(0, e^{2\pi i \zeta })\right|^2
+\left|\theta_2(0, e^{2\pi i \zeta })\right|^2\right\}
    \label{beta}   \end{eqnarray} characterizes dependence on lattice parameters $a$, $b$,
and $\alpha$ via the complex parameter 
\begin{equation}
\zeta ={b\over a}e^{i\alpha}=\rho+i \sigma~.
       \end{equation} Here 
\begin{eqnarray}
\theta_2(z, q )=\sum_{n=-\infty}^\infty q^{(n+1/2)^2} \cos (2n+1)z \nonumber \\
\theta_3(z, q )=\sum_{n=-\infty}^\infty  q^{n^2} \cos 2nz  
       \end{eqnarray} are theta functions \cite{AS}.  The minimum of the interaction energy
correspond to the triangular vortex lattice with $\beta=1.1596$, 
$a=b=2l\sqrt{\pi/\sqrt{3}}$, $\alpha=\pi/3$.  According to Eq. (\ref{GL}) the Gibbs
potential has a minimum  at the particle density $n=(\mu -\hbar \Omega) / \beta g$.  The
vortex density $n_v=1/2\pi l^2$ is equal to the density of LLL states. 

Let us now take into account the parabolic trapping potential. In equilibrium the usual
Thomas-Fermi condition takes place:
\begin{equation}
\mu(r) + m{(\omega_\perp^2 -\Omega^2)r^2 \over 2}=\mu(0)~,
     \label{mu}  \end{equation} where $\mu(0)$ is the chemical potential in the center of
the trap, $r=0$. This gives the inverted parabola for the density distribution in the trap:
\begin{equation} n(r)=n(0)- {m(\omega_\perp^2 -\Omega^2)r^2 \over 2 \beta g}~.
   \label{par}  \end{equation} The condition $n(R)=0$ yields the cloud radius $R=\tilde
c_s(0)\sqrt{2/(\omega_\perp^2 -\Omega^2)}$, where
$\tilde c_s(0) =\sqrt{\beta gn(0)}$ is the sound velocity in the center ($r=0$) of the
rotating BEC. 

The authors of Refs. \cite{WBP,Dal,CKR} also received the inverted-parabola density
profile. In their approach it looks as the effect of  small distortions of the vortex
lattice, which essentially transforms the Gaussian  density profile predicted by Ho
\cite{Ho} for an ideally regular vortex lattice in a trap. But sensitivity of the Gaussian
profile to small distortions is nothing more than evidence that the Gaussian profile is
unstable. Small distortions are inevitable in a finite vortex cluster (see Sec.
V.B  in Ref. \cite{RMP}), but it does not mean that they are crucial for bulk properties of
a macroscopic vortex crystal. Our approach yields the inverted-parabola
density profile without paying attention to small distortions, because the Gaussian
profile does not appear at any stage of the analysis.

Let us consider restrictions on existence of the LLL regime. First, the energy of the
lowest Landau level, $\hbar \Omega$, should exceed the interaction energy per particle,
$\beta gn$ (but $\beta$ is close to unity and is not essential for an order-of-magnitude
estimation). This yields the inequality $n\ll \hbar \Omega /g$, which is equivalent to
the inequalities $\xi \gg b$ or $c_T \gg c_s$ where $c_s =\sqrt{gn}$ and
$c_T=\sqrt{\kappa \Omega/8\pi}$ are the sound and the Tkachenko mode velocity in the VLL
regime. Second, the BEC with a regular vortex lattice exists as far as the filling factor
$n/n_v$ (the number of particles per vortex) exceeds  unity (see below), i.e., the
inequality $n \gg \Omega/\kappa$ is required. The two inequalities are compatible for a
weakly interacting bose gas with $g\ll h^2/m$. Since $g \sim h^2a_s/m l_z$, it is necessary
that the oscillator length $l_z $ for the trapping potential localizing the BEC cloud along
the rotation axis exceeds the scattering length $a_s$.  One can rewrite these inequalities
in terms of the total number of particles $N=\pi n R^2$. Since $R^2 =2\beta gn
/m(\omega_\perp^2-\Omega^2)$, the order-of-magnitude estimation for density is $n \sim
\sqrt{mN(\omega_\perp^2-\Omega^2)/g}$. Then the LLL regime takes place if
\begin{equation} {\Omega^2 \over \omega_\perp^2-\Omega^2}{ gm \over h^2} \ll N \ll
{\Omega^2 \over \omega_\perp^2-\Omega^2} {h^2\over  gm}~.
     \end{equation} One more condition is the presence of many vortices in the cloud: $\pi
n_v R^2 \gg 1$. This yields the inequality
\begin{equation}
 N \gg {\omega_\perp^2-\Omega^2 \over \Omega^2} {h^2\over gm}~.
     \end{equation} This is compatible with the previous inequalities  only for rapid
rotation when $\omega_\perp^2-\Omega^2 \ll
\Omega^2$. 

Deformation of the triangular lattice should increase its energy. On the basis of the
elasticity theory for a 2D crystal with hexagonal symmetry \cite{LL}, the elastic energy
should be
\begin{equation} E_{el}= C_1 (u_{xx}+u_{yy})^2 +C_2 [(u_{xx}-u_{yy})^2+ 4 u_{xy}^2]\,
   \label{el}    \end{equation} where $C_1$ is the compressional modulus, $C_2=mn\tilde
c_T^2/2$ is the shear modulus of the vortex lattice, 
$\tilde c_T$ is  the Tkachenko wave velocity in the LLL regime, and
$u_{ij}= {1\over 2} (\nabla_i u_j+\nabla_j u_i)$ are components of the deformation tensor.
Since the parameter $\beta$ does not depend on the vortex  density, the compressional
modulus vanishes: $C_1=0$ \cite{GB}.  In order to find the shear modulus let us deform the
triangular lattice as shown  in Fig.
\ref{fig1}. The variation of the complex parameter
$\zeta$ is proportional to the shear deformation: $\delta \zeta =\delta \rho = 2 u_{xy} \sin
\alpha$.  Expanding the expression Eq. (\ref{beta}) for $\beta$ and comparing  the term
$\propto \delta \rho^2$ to the thermodynamic potential, Eq. (\ref{GL}), with the elastic
energy Eq. (\ref{el}), one obtains the value of the shear modulus:
\begin{equation} C_2={gn^2\over 4} {\partial^2 \beta \over \partial \rho^2}\sin
\alpha= 0.1191 gn^2=0.1027 mn
\tilde c_s^2~.
    \label{C-2}   \end{equation}
This exactly agrees with the numerous calculations of the shear modulus
$c_{66}=2C_2$ for the flux lattice in type II superconductors close to the critical field
$H_{c2}$ \cite{sup}. But in this work it was calculated
anew since Eq. (\ref{C-2}) yields the result  2 times larger than the value by
Sinova {et al} \cite{SHM} and 10 times larger than the value $(81/80 \pi^4) mn\tilde
c_s^2$ received by Baym
\cite{Ba}. This numerical difference is important for interpretation of the
experiment (see below).

One can proceed with the analysis of the Tkachenko-mode spectrum on the LLL regime
using the same hydrodynamic equations as in the VLL regime \cite{RMP,Baym,ST}, but
with $c_s$ and $c_T$ replaced by  $\tilde c_s$ and $\tilde c_T$. However, since the LLL
regime is possible only for very rapid rotation when
$\omega_\perp- \Omega \ll \omega_\perp$ the effect of high BEC compressibility is always
strong (see the comment [16] in Ref. \cite{GB}) transforming the Tkachenko mode spectrum
to  quadratic:
$\omega=\tilde c_s\tilde c_Tk^2 /2\Omega$. In this limit one can use the analytic
expression for the Tkachenko eigenmodes of a finite BEC cloud in a parabolic trap derived
for the VLL regime in  Ref. \onlinecite{ST}: $\tilde \omega_i =\gamma_i /s$. Here $\tilde
\omega_i$ is the value of the reduced frequency $\tilde \omega =\sqrt{\sqrt{3}/8\pi}\omega
R/c_T$ and $s= 2\sqrt{2}\Omega/\sqrt{\omega_\perp^2 -\Omega^2}$. The numbers $\gamma_i$
depend on the number $i$ of the eigenmode. The two lowest eigenmodes correspond to
$\gamma_1= 7.17$ and $\gamma_2=16.9$. Using the expression 
$R=\sqrt{2}c_s(0)/\sqrt{\omega_\perp^2 -\Omega^2}$ for the cloud radius one obtains 
$\omega_i= 
\sqrt{\pi/\sqrt{3}} \gamma_i (c_T/c_s) (\omega_\perp^2 -\Omega^2)/\Omega \approx  
2\sqrt{\pi/\sqrt{3}} \gamma_i (c_T/c_s) (\omega_\perp -\Omega)$. Applying this expression
to the LLL regime one should replace $c_s$ and $c_T$ with $\tilde c_s =\sqrt{\beta} c_s$
and $\tilde c_T$ given by Eq.(\ref{C-2}). This yields the Tkachenko eigenfrequencies
$\omega_i \approx 0.8633 \gamma_i (\omega_\perp -\Omega)$. The experiments on rapid
rotation \cite{Cod} roughly agree with expected linear dependence of the first Tkachenko
mode on
$\omega_\perp -\Omega$, but the slope of the experimental line is about 4 times less than
our prediction for the LLL regime. It could be an evidence that the  experiment has not yet
reached the LLL limit. Since experimental values of $(\omega_\perp-\Omega)/\omega_\perp$
look small enough, apparently in order to approach to the LLL limit further, the experiment
should be done with a smaller number of atoms.

What should happens with the LLL regime when the filling factor $n/n_v$ approaches unity?
Possible answers to this question are now intensively studied by theoreticians numerically
and analytically
\cite{SHM,CWG,Ba}. The  following simple discussion certainly cannot be a substitute of
these investigations, but hopefully would be useful for better qualitative understanding of
possible scenarios. At low 
$n/n_v$ bose condensation is destroyed and  a single macroscopic wave function cannot
describe all atoms anymore
\cite{SHM}. For an illustration let us construct a state of the bose system, which
successfully competes with the BEC state at low filling factors. The single-particle state
\begin{equation}
\psi_i= {1\over \sqrt{2\pi} l} \exp\left[-{(x-X_i)^2+(y-Y_i)^2\over 4l^2}\right]
      \label{lar}  \end{equation} corresponds to the classical Larmor orbit of a charged
particle centered in the point with coordinates
$(X_i,Y_i)$. It is a solution of Eq. (\ref{Sch}) for the gauge $\vec v_0(-\Omega y, \Omega
x)$ and  belongs to the space of the LLL states. Let us assume that any LLL state given by
Eq. (\ref{lar}) cannot have more than one atom (the number $N$ of them is much less than
the number of LLL states). Then the wave function of $N$ bosons can be written as 
\begin{equation}
\Psi={1\over N!} \sum _{P} \prod _{i=1}^N \psi_i (x_i, y_i)~.
        \end{equation} Here $i$ is the number of the particle (subscript of $x$ and $y$)
and the number of the occupied site ($X_i,Y_i$) (subscript of
$\psi$). In  contrast to BEC, all atoms are in different states, and proper symmetrization
(summation over all permutations $P$) should be done. Strictly speaking the states $\psi_i$
given by Eq. (\ref{lar}) are not orthogonal. Therefore the wave function
$\Psi$ requires some normalization factor. But for small filling factors overlapping of the
states is weak and an additional normalization factor is exponentially close to unity. The
next step is to calculate the interaction energy for this state, Only interaction between
closest neighbors is essential. The cross interaction term between them is 
\begin{equation} {g\over 2} \int dx\,dy\,|\psi_i|^2|\psi_{i+1}|^2={g\over 8\pi l^2}
e^{-4r_0^2/l^2}~.
        \end{equation} Here $r_0$ is the distance between two sites. Assuming the
triangular-lattice ordering of occupied sites,  $r_0 =\sqrt{2/\sqrt{3}n}$. Collecting all
terms of this value we receive for the total interaction energy integrated over the whole
space:
\begin{equation} E_{cr} ={6N N!\over N!}{g\over 8\pi l^2} e^{-4r_0^2/l^2} ={3 \over
2}gNn_ve^{-\pi n_v  /\sqrt{3}n}~.
     \label{cr}   \end{equation} Though we used the properly symmetrized wave function of
$N$ bosons, in fact statistics is not important and the same energy can be derived for an
unsymmetrized function.   Comparing the energy Eq. (\ref{cr}) with the total energy of the
BEC state,
$E_B ={1\over 2}g\beta Nn$,  one sees that at small filling factors $n/n_v \ll 1$ the BEC
state has a larger energy. The transition between two states is expected at $n/n_v \sim 1$,
but the present calculation becomes inaccurate there. This estimation certainly cannot
pretend to be an evidence that some crystal structure appears at small $n/n_v$, but it
demonstrates that in this limit the BEC is not the equilibrium state. 

Another option at $n/n_v \sim 1$ is vortex melting without destruction of BE condensation 
\cite{SHM,CWG,Ba}. The plane Tkachenko mode
$u(k) e^{ikx -i\omega t}$, where $u(k)$ is the amplitude of vortex displacements, has the
energy $\sim L^2 u(k)^2 mn
\tilde c_T^2k^2$ in the square $L\times L$. Considering the quantum melting at zero
temperature this energy is equal to $\hbar \omega(k)/2$. The average vortex displacement
squared is obtained by integration of $u(k)^2$ over the Brillouin zone of the vortex 
lattice:
\begin{eqnarray}
\langle u^2\rangle \sim L^2\int_0^{1/b} k\,dk\,u(k)^2 \sim \int_0^{1/b} k\,dk\,{\hbar
\omega(k) \over mn \tilde c_T^2k^2} \sim {1\over n} 
        \end{eqnarray} According to the Lindemann criterion, quantum melting is expected,
when $\sqrt{\langle u^2\rangle}$ approaches to the intervortex distance $b \sim 1/\sqrt{n_v}$.
This corresponds to the filling factor $n/n_v$  of order unity. So vortex-lattice melting
can precede destruction of the BEC. 

In summary, the ground state of a rapidly rotating Bose-Einstein condensate is analyzed in
the LLL regime using the exact wave function known for type II superconductors close to the
upper critical magnetic field $H_{c2}$. The analysis yields the inverted-parabola
density distribution in a parabolic trap and  the exact value of the shear elastic
modulus of the vortex lattice in the LLL regime, which exceeds the values received in the
previous calculations for BEC. This has an impact on interpretation of recent experiments
on very rapid rotating BEC.

I thank Alexander Fetter for interesting stimulating discussions of the LLL regime.
Ernest-Helmut Brandt helped a lot to my search of literature on vortex elasticity in type II
superconductors.

\end{document}